\documentclass{article}

\usepackage[english]{babel}

\usepackage[letterpaper,top=2cm,bottom=2cm,left=3cm,right=3cm,marginparwidth=1.75cm]{geometry}

\usepackage{amsmath}
\usepackage{graphicx}
\usepackage{titlesec}
\usepackage[colorlinks=true, allcolors=blue]{hyperref}

\makeatletter
\titleformat{\subsection}[hang]{\large\bfseries}{Question \thesection.\@arabic\c@subsection\hspace{2mm}}{0pt}{\slshape}{}
\makeatother

\title{AI (r)evolution – where are we heading?\\
Thoughts about the future of music and sound technologies in the era of deep learning}
\author{Authors: %\textcolor{green}{(students in list ordered alphabetically by surname with affiliation either $^1$ or $^2$}\\ \textcolor{green}{please add your name here and use line break when too many names forhttps://www.overleaf.com/project/63ee659081f396089aba82bc one line)}\\
Giovanni Bindi$^1$,
Nils Demerlé$^1$,
Rodrigo Diaz$^2$,
David Genova$^1$,\\
Aliénor Golvet$^1$,
Ben Hayes$^2$,
Jiawen Huang$^2$,
Lele Liu$^2$,\\
Vincent Martos$^1$,
Sarah Nabi$^1$,
Teresa Pelinski$^2$,
Lenny Renault$^1$,\\
Saurjya Sarkar$^2$,
Pedro Sarmento$^2$,
Cyrus Vahidi $^2$,
Lewis Wolstanholme\(^2\),\\
Yixiao Zhang$^2$
\\
\\
Editors: Axel Roebel$^1$, Nick Bryan-Kinns$^2$$^*$, Jean-Louis Giavitto$^1$, and Mathieu Barthet$^2$}
\date{%
    $^1$UMR 9912 STMS -- IRCAM, CNRS, Sorbonne University, French Ministry of Culture, Paris, France\\%
    $^2$UKRI Centre for Doctoral Training in Artificial Intelligence and Music (AIM)\\
    Queen Mary University of London, UK\\[2ex]%
    \today
}

\begin{document}
\maketitle

\begin{abstract}
Artificial Intelligence (AI) technologies such as deep learning are evolving very quickly bringing many changes to our everyday lives. To explore the future impact and potential of AI in the field of music and sound technologies a doctoral day was held between Queen Mary University of London (QMUL, UK) and Sciences et Technologies de la Musique et du Son (STMS, France). Prompt questions about current trends in AI and music were generated by academics from QMUL and STMS. Students from the two institutions then debated these questions. This report presents a summary of the student debates on the topics of: Data, Impact, and the Environment; Responsible Innovation and Creative Practice; Creativity and Bias; and From Tools to the Singularity. The students represent the future generation of AI and music researchers. The academics represent the incumbent establishment. The student debates reported here capture visions, dreams, concerns, uncertainties, and contentious issues for the future of AI and music as the establishment is rightfully challenged by the next generation.
\end{abstract}

\section{Introduction}

Deep learning-based technologies are evolving very quickly, and seem to become the basis of numerous changes in our everyday life. However, self-driving cars, conversational agents, machine-based language translation, and image generators are only the tip of the iceberg. Machine learning techniques are developed for a growing list of application domains: medicine, smart cities, humanoid robots, management of electricity grids, and they are starting to be used in fundamental research disciplines like physics, chemistry, genetics, and even mathematics.

Given the apparent proliferation of technologies that enable learning from data, the Queen Mary University of London (QMUL) \& Sciences et Technologies de la Musique et du Son (STMS) doctoral day held on 13 Feb 2023 aimed to stimulate discussions about the future impact and potential of Artificial Intelligence (AI) in the field of music and sound technologies. Eight prompt questions about current trends in AI and music, and discourse surrounding AI were generated by academics from QMUL and STMS. Students from the two institutions then formed teams to discuss and debate these questions. This report presents a summary of the discussions by research students from QMUL and STMS. Given the healthy debate around the questions, it must be noted that the opinions in this text are not necessarily shared by all authors.

\section{Data, Impact, and the Environment (group A)}
%Group A

\subsection{Machine learning systems rely on large amounts of data that are difficult and expensive to generate and process. How can we ensure that academic research remains competitive and innovative compared to the various industrial players that have access to better computational resources and larger amounts of data?}

Firstly, redundant effort on the same task could be avoided: once there is enough industrial interest in a specific task, it could be a sign for the academia to shift their focus towards unsolved research questions. For example, in the field of source separation, the first breakthroughs happened with \cite{uhlich2017improving} and U-net \cite{DBLP:conf/ismir/JanssonHMBKW17} came out in 2017 by Sony and Spotify respectively, but based on private datasets. After a few years of the SiSEC (Signal Separation Evaluation Campaign) challenge running and more public datasets made available did we finally come across an open-source implementation, Open-Unmix \cite{stoter2019open}, that matched industry model performance while using the same architecture from \cite{uhlich2017improving}. Subsequently, Deezer released Spleeter \cite{spleeter2020}, using a very similar U-net architecture. While the contribution may be significant in terms of making tools more accessible to a wider audience, it appears to be limited in terms of methodology. Since then, the state-of-the-art for music source separation has consistently been pushed by industry research, such as the latest Hybrid Demucs model \cite{defossez2021hybrid}, which often relies on smaller contributions from academic researchers to improve performance. At this point, we see a trend of models trained by academia (such as LaSAFTNet \cite{choi2021lasaft}) on publicly available datasets that rarely outperform state-of-the-art models released by industry giants under the common evaluation setting. 
While their contributions may not always lead to major breakthroughs in the field, they are nevertheless crucial in advancing research in specific cases uncovered by the industry such as instrument-specific separation, acappella separation, blind separation, etc.
% Meanwhile their contributions are extremely crucial to improving the field, but the prestige and the final beneficiaries are always the industry players who are able to simultaneously keep their edge while benefiting from publicly funded research.

Secondly, it is crucial to encourage the creation of finished prototypes or ready-to-use tools in academia, properly licensed to avoid situations where a company reimplements it and sells it as a commercial product. This is difficult for some products, such as hardware. And it requires efforts to maintain a product. Therefore ideally industry and academia could work together in developing and maintaining a product, through technology transfer, including licensing and spinout. To this end, academia should recognize accomplishment in releasing finished prototypes so that student researchers would be more encouraged to do so.

Thirdly, while huge models like GPTs are trained to solve the general problem of natural language understanding and generation, academic researchers could aim at solving specific research questions with smaller models and limited resources. Academia could take advantage of the large pretrained models from the industry, and the industry could get insights on improving every single task. This approach could lead to more efficient use of resources and a better understanding of how to solve specific research questions.

In conclusion, by avoiding redundant efforts, encouraging creating ready-to-use tools and technology transfer, and utilizing pre-trained models from the industry, academia can remain competitive and innovative despite the better resources and larger datasets available to industrial players.

\subsection{What can be done about the environmental impact of deep learning approaches?}

The increasing use of AI in academic research and the industry is raising concerns about its environmental impact. Several practices have been suggested in \cite{strubell2019energy}, including reporting training time, sensitivity for machine learning models, and sharing local infrastructure. A set of efficiency measures is proposed in \cite{schwartz2020green}, such as carbon emission and electricity usage.

In this context, there are other initiatives that can be taken to address these concerns.
First, researchers should be encouraged to report the amount of energy consumption or carbon emissions used in their publications. This information will help raise awareness of the environmental impact of AI research, and incentivize researchers to develop more efficient algorithms and training methods.

Secondly, a task-specific energy rating system can be implemented to label the energy level of open-source pretrained models. This rating system could be displayed on open source model hubs such as huggingface \footnote{https://huggingface.co/}. By doing so, people can get an idea of the energy consumption behind each model, and choose a more energy-efficient model for their specific task.

However, it is important to note that the more efficient a model becomes, the more it would be used, leading to equal or more energy consumption. This paradoxical situation is similar to the development of engines. The overall increase in energy consumption seems to be inevitable.

Back in 2009, there was a debatable article saying that "Performing two Google searches from a desk computer can generate about the same amount of carbon dioxide as boiling a kettle for a cup of tea" \footnote{https://searchengineland.com/calculating-the-carbon-footprint-of-a-google-search-16105}. Another evolution with ChatGPT by Microsoft is happening right now. Search engines have become a big part of people’s lives, and we cannot abandon them. Similarly, AI is necessary for the advancement of technology, but we should be concerned about energy consumption and continue to explore ways to reduce it.

Meanwhile, we can be optimistic about the use of eco-friendly energy sources such as solar and wind energy, as well as more energy-efficient computers such as biological computers becoming more accessible in the near future.

%Group B
\section{Responsible Innovation and Creative Practice (group B)}
\subsection{What does responsible innovation mean for the field of AI for sound and music? How should we envision the use of AI in sound processing and music in a way that artists would find rewarding, and how can we avoid negative impacts, for example for composers and performers?}

In recent years, we have observed a considerable rise in public interest towards AI generative models. Many discussions have arisen in the media which question whether image generation tools will eventually `replace' visual artists, alongside assessing the ethics of using artists' work to train these models without their economic compensation \cite{santosCanAIGeneratedArt2022, thompsonAIArtGenerators2023}. Recent advancements in audio generation have extended these ethical discussions to the music industry, where musicians are now also beginning to face very similar issues. 

Whilst there are many commonalities between the domains of the visual arts and music, such as models of remittance/retribution, it appears that the bond between spectator and artist is generally stronger in popular music than in the visual arts. With these economic concerns in mind, it is hard to imagine a fan culture centered around AI generated music, at least in the form we see today towards our most popular artists. A more immediate issue, however, is that of AI technologies automating the jobs, such as sound engineering or mastering, which many musicians undertake to support themselves financially alongside their composition and performance work. The rate at which new AI models are released calls for a discussion across disciplines with regards to how technology's impact upon these professions may be mitigated. We suggest that, rather than developing tools that aim to \textit{solve} a problem (e.g., solve automatic mixing), we should instead aim to create tools that \textit{assist} or \textit{support} the artist's work in that particular task \cite{ozmengaribaySixHumancenteredArtificial2023}. Ultimately, the AI technologies (i.e., the models) would be the same, but instead of being presented as close-ended generators, they could be developed as interfaces musicians can interact with and include in their workflows. Even the best automatic mixing model does not possess the same reflexivity that an expert mixing engineer has - a sound engineer might have listened to a particular piece of music earlier on in the day that inspires them to approach mixing in a new way. Such a dynamicity has not yet been encountered in our current AI music models, which are yet to adopt such an approach to their materials and past predictions.

The rigidity of these AI models has further implications in how we define and understand our musical idioms. Typically, AI technologies are designed to both codify patterns for use in generative environments and extract dominant characteristics from arbitrary data points.  
The prominence of these patterns and characteristics are typically influenced through a curated training and evaluation process which, when employed for creative purposes, serve to imbue some desired semantic/semiotic/qualitative values. 
In the sound and music domain, these characteristics are often inferred from the tradition of Western music theory, particularly functional harmony, metric understandings of rhythm, standardised instrumental/orchestral forces and the cultural artefacts that symbolically coincide with them.
Although Western music theory has it uses, and is both studied and understood by many, its effectiveness in both practice and technological contexts is intrinsically limited.
In many cases, music theory does not even apply to many Western music contexts, and it also struggles to remain relevant when extended to other cultural traditions (see e.g. \cite{agawuRepresentingAfricanMusic1992}).
As a result of this, practitioners tend to have a dynamic relationship towards their understanding of music theory, employing it instinctively, and defying it or redefining it on a regular basis.
These limitations of Western music theory are similarly transferred to the technologies that robustly and naively center around them, which detracts from our more general cultural understanding as much as it limits the applicability and scope of these technologies.

Moreover, the development of AI technology, as it is currently organised, leaves little place for critical discussion on `why' and `how' our AI tools are created outside of a race for scientific progress. AI technologies are too often immediately deployed and shared as functional and effective tools without concern for the ethical and sociotechnical questions that the concept of a `tool' entails. This is heightened by the frequent use of analogies between neural networks and biological processes in both research and pedagogical contexts (e.g. the analogy between the brain neuron and the neuron from a neural network). This drives a narrative in which AI technologies are `naturalised', giving them a sort of autonomy, as if AI researchers and developers' tasks were to discover or to unearth the natural processes behind the `already existing' AI and deep learning technologies, akin to studying animal or human intelligence. % \cite{Rahwan2019}. 
Instead, it must be stated that these technologies are human-made, and that society, and (especially) researchers and programmers, hold responsibility for how and why they are created, the biases they contain and the position they occupy within our culture.

A de-naturalisation or de-mystification of AI would contribute to refocusing responsibility towards the researchers and developers that build these technologies. That being said, the impact of these technologies is not foreseeable from the perspective of a single discipline.
Responsibility is achieved through interdisciplinary research narratives - a responsible innovation takes care over all disciplines which relate to the specific topic of study.
What Donna Haraway termed ``response-ability''~\cite{harawayStayingTroubleMaking2016}, Debaise \& Stengers describe as ``the capacity to be accountable for an action or an idea to those for whom the action or idea will have consequences''~\cite[p.17]{debaiseInsistencePossiblesSpeculative2017}.
Response-ability, in this sense, is the desire for technological research and culture to work together to preserve and strengthen their interdependence. 
It encourages them to come ``face-to-face''~\cite{harawayCompanionSpeciesManifesto2003} with one another, to be able to empathise with one another, and to remain considerate and explicit with regards to the power that one might have over the other.
As technological and cultural landscapes evolve sympathetically, it is the responsibility of those who forge them to engender and maintain their togetherness.

\subsection{Currently, most research activities dealing with AI for sound and music focus around generation, analysis, and synthesis. Can we imagine AI contributions to music playing, music listening, musical performances, manufacturing of (augmented) instruments, acoustic contexts such as concert halls, the diffusion or design of sound, or other domains not yet considered?}

Mainstream media, industries and publications tend to gravitate towards AI models that focus upon tasks such as music generation conditioned on artists or genres (e.g., Jukebox \cite{dhariwalJukeboxGenerativeModel2020}) or, more recently, text-to-music generation (e.g., MusicLM \cite{agostinelliMusicLMGeneratingMusic2023}). These tasks are highlighted and celebrated for their impressive coverage and holistic scope, whilst the achievements of many smaller AI technologies are often easy to understate and overlook. More situated, ambiguous or highly contextualised uses of AI, such as augmented instrument design, do not receive the same mainstream attention, despite their own ingenuity and prowess. The more specific an application of AI becomes, or the more situated it becomes within a particular art practice, the more likely it is to occupy a niche position in the overall conversation surrounding AI for sound and music.

The prevalent tasks in sound and music generation have also carved out a particular space in academia, encouraging the formation of standardised benchmarks, the establishment of objective evaluation metrics, and the creation of publications which focus solely on the evaluation of these AI models~\cite{kilgourFrechetAudioDistance2019}. Tasks that are harder to evaluate or standardise have greater difficulty entering into the academic publication schema with the same status. 
As these larger generative tasks are typically more open–ended and relatable, there is much more opportunity for others to become influenced by these models, and continually develop upon them.
This encourages research to fall into distinct trends, where technologies linearly influence the development of one another, and tasks become more generalised as they develop.
As a result, agreed upon quantitative standards arise to compare the multitude of approaches towards a given task, and the breadth and scope of these evaluations is effectively narrowed.
This progress narrative, akin to what Thomas Kuhn describes as ``normal science''~\cite{kuhnStructureScientificRevolutions1996}, produces the perception that everybody is working on the same thing, and marginalises those who are working outside of these dominant approaches.

As the development and use cases of these mainstream technologies become more generalised, they too begin to have subversive effects on cultural practice and its related industries.
Technological pursuits which may have originated from ideas relating to the creation of sounds in specific contexts, have now matured and become generalised synthesis tools, equally applicable to a wide range of scenarios. 
And as these tools become more widely applicable, they also obfuscate the need for more context-dependent devices, as well as techniques for some of our more underused and unique musical practices and idioms.
In line with this idea of `normal science', we can think of these technologies as encouraging a sense of `normal practice', whereby cultural development is fixated on perfecting some major aspect of its activity, as opposed to encouraging experimentalism and the accumulation of new practice driven techniques. 
As far as the grander narrative of computer music research is concerned, this is a distinct shift from the practices of our previous cultures of research - those whom, like Jean Claude-Risset and Miller Puckette, engaged in their research as means of furthering both their own creative practice and the landscape of potential creative techniques and ideas \cite{rissetComputerMusicWhy2003}.

In terms of responsible approaches towards AI, and the development of technologies that are successful outside the dominant trends of generative music, synthesis and analysis, it is important to remain situated within and connected to exploratory and experimental arts practices.
As an incentive for research, many of the fringe aesthetics involved in these cultures incite the curation of new directives and techniques for creative and technological expression. 
Although these practices generally fall outside of the mainstream, and similarly will not receive the same media and industrial attention as the larger AI models do, they are just as important for the development of technology as they are for the continued growth of our cultural narrative and understandings. 
Where the ethical quandaries towards artists and their practices are concerned, research that pertains a strong relationship with the arts and its development aligns itself with Haraway's aforementioned idea of ``response-ability''~\cite{harawayStayingTroubleMaking2016}.
In this sense, technological development and research may embody the same sense of creativity and inventiveness as the cultural practices it works alongside.
And in doing so, in supporting these more situated affinities, the overarching narrative of progress is mitigated and directed away from a culture of normal practice and obfuscation, and towards one where artists, musicians and technologists can continue to prosper together, in sympathy and interdependently.

\section{Creativity and Bias (group C)}
%Group C
\subsection{AI generators like DALL·E 2, or MusicLM produce media content based on text prompts, but can we call these systems creative? What is creativity and what are the characteristics of an activity we would want to call creative? To what extent can we expect to find these characteristics in a DNN?}

% - the models are very good at imitating human creativity, which we believe is not the same as being creative\\
% - creativity demands intention (intentional creative act), and DNNs don't have intentions -> shift the focus on modeling the creative process rather then the outcome\\
% - however, the models can be used, by humans\\
% - often, works where the users/artists try to "break" the system are the most "creative"
% - creativity is an iterative/colaborative/exploratory process\\
% - surprise \\
% - scaling to large datasets leads to loss of personalization (?)

Prior to assessing the creative potential of deep generative models such as DALL·E \cite{ramesh2022hierarchical} or MusicLM \cite{agostinelli2023musiclm}, one should discuss the core concept of \textit{creativity} itself, which is complex and challenging to define precisely. Among the many attempts to define creativity, one can identify recurrent properties which are shared by the majority of literature on the topic, namely the ideas of \textit{novelty}, \textit{intention} and \textit{cultural relevance} \cite{esling2020creativity}. Creative acts are purposeful and deliberate, involving the author's intent and choices, which is fundamentally different from the process of imitation. As current deep generative models like DALL·E and MusicLM are trained to extract statistical patterns from the training data, they lack the required intentionality to pursue a true creative act. Furthermore, the authors in \cite{penhaWillMachinicArt} argue that, even in a scenario where machines would develop a sense of volition/intention, thus being able to create something uninfluenced by human-made art, it would not be possible for us to understand their creative outcome, for we are bound to frame such outcome from a human perspective.

Additionally, we suggest that the best we can aim at, as human beings standing by an artistic creation by another entity, is to an understanding of what could have motivated another human being to create such a work. %
%As such, we should not be able to understand an artistic creation originating by an artificial mind with a physical experience of the world that differs from our own, even if they have a privileged access to our culture. The boundaries for this incomprehension are those of the human mind.

These systems can also be intended as proposal generators, which can be refined by the end user. One interesting aspect is that many artists deliberately deviate from the original objective of such models by feeding them unexpected inputs. To this extent, they seek to break the initial system in order to fully leverage all the expressive abilities of these models \cite{gold30217}.

In the end, we believe that generative systems should not be aimed at replacing humans but rather as offering co-creative tools \cite{esling2020creativity} which could complement and extend current instruments and they should be designed to reflect the will to pursue artistic intentions.

\subsection{How can we reduce the bias in machine learning models? E.g. bias towards Western musical cultures?}

% - diversity encouraging tracks on MIR conferences (e.g. ISMIR 2022 special track)\\
% - creation of non-western music datasets \\
% - biases are sometimes task-oriented/useful (e.g. user-oriented DL products, in which biases from a user perspective are desirable) + inductive biases in deep models that ease down the learning process \\

There are many aspects to be considered when reflecting on biases in machine learning models. It is important to recognize that we cannot completely unbias the output of a generative model, since there will always be an unremovable source of bias coming from the data used to train the system on. Still, cultural biases can represent a discriminating factor towards under-represented scientific and artistic communities. When attempting to reduce such components, encouraging diversity should be the main goal. To this end, we believe that it is important to reinforce the need for datasets that target culturally different segments. The creation of special conference tracks or journal special issues encouraging the production and release of datasets that are culturally diverse (e.g. International Society for Music Information Retrieval (ISMIR)) can help stimulating the development of models that are not biased towards, e.g., Western musical cultures, bringing together researchers and practitioners from different backgrounds and cultures.

It is also important to recognize that biases can sometimes be task-oriented or useful, especially in user-oriented deep learning products where biases from a user perspective may be desirable. For example, a recommendation system for a particular genre of music may be biased towards that genre, but it may also be useful for users who are looking for music in that genre. It is essential to evaluate biases on a case-by-case basis to determine whether they are helpful or detrimental. % unsure

Finally, introducing the correct inductive biases in deep models can help ease down the learning process and reduce the number of training examples required \cite{goyal2022inductive}. It is thus important to carefully differentiate whether bias can potentially be harmful, e.g. coming from a strongly biased dataset, or useful, as in the case where modelling choices can enhance the learning capabilities of such artificial agents.

\section{From Tools to the Singularity (group D)}
%Group D

\subsection{Is the singularity around the next corner or are there implicit limitations to AI systems that will keep them ``dumb'' irrespective of the amount of data we train them with?}

The concept of singularity - a point in time when artificial intelligence surpasses human intelligence and becomes self-improving \cite{eden2012singularity} - has been the subject of much speculation and debate. In our discussion, we explored the question of whether the singularity is around the next corner or if there are inherent limitations that will keep AI systems ``dumb'' regardless of the amount of data used to train them.

While we did not arrive at a definitive answer, we acknowledged that there have been numerous breakthroughs in AI recently that are becoming more frequent. Early in 2017, R-net \cite{wang2017r} has demonstrated performance beyond the level of human experts in the Question Answering task; Following the theory of scaling law \cite{kaplan2020scaling}, recent large models containing massive parameters and trained with massive data have indeed shown better performance than previous AI models on various tasks \cite{kocon2023chatgpt}. 
% Yixiao: I added some statements in the paragraph above.

% 1. Ethics problems.

There are many challenges and limitations that AI systems face, even with large amounts of data. The first thing that should be considered is the potential ethical issues. For instance, there are concerns around data privacy, stereotypes, and the ethical implications of AI-driven decision-making \cite{bommasani2021ethical, kolesnyk2021large, hovy2020stereotyping}. These issues are already being manifested in a wide range of AI applications and have been embodied as never before with the development of large-scale models, from predictive policing and facial recognition technology \cite{kitchin2019ethical}, to disputes over copyright ownership \cite{mittelstadt2021ethics}.

% Yixiao: I rewrite this para.
In this context, the intelligence exhibited by AI systems may be criticized as "dumb" since it cannot guarantee responsible output. Given the rapidly evolving nature of these technologies, it is essential for humans to consider the ethical implications of AI systems and engage in ongoing dialogues to best manage their associated risks and benefits. However, it is challenging to address this issue fundamentally due to the neural network structure of large AI models, which often lack interpretability \cite{xu2019explainable}. Despite this challenge, it is crucial to prioritize transparency, accountability, and openness in AI development and engage with a broad range of stakeholders, including government regulators, civil society organizations, and affected communities.% \cite{ozmen2023six}.

% Despite the fascinating nature of the singularity and its potential implications, we suggested that there are more pressing issues related to AI applications that should be given priority. 

% Yixiao: I do not think it is scientific enough...
% Another critical aspect of the singularity that we discussed is the ability for AI to interact with the physical world. While this is seen as a necessary component for the singularity to occur, it is still unclear whether this ability alone is sufficient for AI to become truly intelligent. Some argue that the development of self-awareness and consciousness is also required for true intelligence to emerge, while others contend that these qualities may be unnecessary.

% Yixiao: revised the conclusion.
Overall, our discussion implies that despite the potential for large models to surpass human performance and reach the singularity, they may still exhibit ``dumb'' aspects that need to be addressed. Thus, it is crucial to focus on the more immediate and pressing concerns related to AI applications to ensure that these technologies are developed and deployed in a manner that aligns with our values and promotes the greater good.

\subsection{Do AI systems escape the notion of being tools? Applied to the field of sound and music, are they qualitatively different from digital audio tools? Or is their contribution limited to that of a tool in the hands of artists or listeners, where these new AI-based techniques simply allow doing things better/differently?}

% Yixiao: added an overall para
To answer the question of whether AI systems in sound and music can be seen as more than just tools, it is crucial to first distinguish between a tool and an agent and provide some definitions. Generally, from a philosophical point of view, one could ask whether AI systems have agency, autonomy, intentionality, consciousness, or moral responsibility \cite{list2021group}; from a practical point of view, one could ask whether AI systems can perform tasks that are beyond the capabilities or expectations of human users, or whether they can influence or interact with human users in ways that are not predetermined by their design. In the field of sound and music, we should consider how AI systems affect the roles and relationships between composers, performers, listeners, and critics. 

While music AI systems still operate within the parameters set by their programmers and users, they do have the ability to introduce new levels of surprise, unpredictability, and creativity into the creative process:

\begin{enumerate}
    \item One example of this is the creation of AI-generated music \cite{briot2017deep}, where AI systems can generate original pieces of music based on existing data or specifications. This can be seen as going beyond the role of a tool and into the realm of a creative collaborator or even composer. However, while AI-generated music has been around for at least 20 to 40 years, there has yet to be a sizable community of people coalescing around it, unlike the case with the development of new musical instruments or algorithms in the past.
    \item Another example is the use of AI for sound design \cite{zhou2018visual}, where AI systems can be used to generate or manipulate sounds in ways that would be difficult or impossible for a human to do manually. This can be seen as expanding the creative possibilities of sound design beyond what traditional tools allow.
\end{enumerate}

AI is also playing a larger role in shaping how people consume music through the rise of AI-generated playlists \cite{ferraro2021melon} and personalized music recommendations \cite{fessahaye2019t}. This can be seen as going beyond the role of a tool in the hands of listeners and into the realm of a new type of music curator or even a tastemaker.

However, there are still valid arguments for viewing AI systems in sound and music as simply tools that assist artists and listeners in achieving their goals. It is important to consider the limitations and challenges of these technologies in creating truly surprising and innovative works. Overall, the potential for AI to develop intentionality and generate new forms of art is intriguing, but it also highlights the need for ongoing discussion and exploration in this field.

% Yixiao: the answer is wonderful. I cannot do more refinement.

\section*{Acknowledgments}
The doctoral day was supported by Sciences et Technologies de la Musique et du Son (STMS, France) and by the UKRI Centre for Doctoral Training in Artificial Intelligence and Music, supported jointly by UK Research and Innovation [grant number EP/S022694/1] and Queen Mary University of London (QMUL, UK).

$^*$Nick Bryan-Kinns is now at the Creative Computing Institute, University of the Arts London.

\bibliographystyle{alpha}
\bibliography{sample}

\end{document}